\documentclass[12pt]{article}

\newcommand{\m}{\mathrm}
\newcommand{\be}{\begin{equation}}
\newcommand{\ee}{\end{equation}}
\newcommand{\ba}{\begin{eqnarray}}
\newcommand{\ea}{\end{eqnarray}}

\usepackage{graphicx}
\usepackage{amssymb}
\usepackage{amsmath}
\usepackage[T1]{fontenc} 
\usepackage[ansinew]{inputenc} 
\usepackage[nosort]{cite}
\newcommand{\inbar}{\vrule height1.57ex width.4pt depth0pt}
\newcommand{\SW}{\relax{\hbox{$\ \inbar\kern-.285em{\rm S}$}}}

\begin{document}
\thispagestyle{empty}
\begin{center}

\null \vskip-1truecm \vskip2truecm

{\Large{\bf \textsf{Reconciling Inflation with Anisotropies at Small Scales}}}

{\large{\bf \textsf{}}}

{\large{\bf \textsf{}}}

\vskip1truecm

{\large \textsf{Brett McInnes}}

\vskip1truecm

\textsf{\\  National
  University of Singapore}

\textsf{email: matmcinn@nus.edu.sg}\\

\end{center}
\vskip1truecm \centerline{\textsf{ABSTRACT}} \baselineskip=15pt
\medskip

There have been persistent suggestions, based on several diverse data sets, that some aspects of our Universe were or are not isotropic. It is not easy to develop a coherent theoretical account of such a ``statistical anisotropy'', for, in standard General Relativity, intuition suggests that it contradicts the predictions of the very successful Inflationary hypothesis. We put this intuition on a firm basis, by proving that if we [a] make use of an Inflationary theory in which Inflation isotropises spatial geometry ---$\,$ this, of course, includes the great majority of such theories ---$\,$ and if [b] we insist on assuming that spacetime has a strictly metric geometry (one in which the geometry is completely determined by a metric tensor), then indeed all aspects of the ``Hubble field'' must be isotropic. Conversely, should a statistical anisotropy be confirmed, then either we must contrive to build anisotropy into Inflation from the outset, or we will have to accept that spacetime geometry is not strictly metric. We argue that the second option, implemented by allowing spacetime \emph{torsion} to be non-zero, would be by far the most natural way to accommodate \emph{some} such observations. Such theories can reconcile a non-isotropic Hubble field at the end of Inflation with a perfectly isotropic spatial geometry, and thus are able to reconcile Inflation with certain kinds of possibly observable anisotropies at relatively small cosmic scales (though by no means with all).

\newpage
\addtocounter{section}{1}
\section* {\large{\textsf{1. Inflation and the Hubble Field }}}
Cosmic Inflation \cite{kn:inflationaris,kn:linde} gives a convincing account of the formation of large-scale cosmic structure and explains the observed almost scale-invariant spectra of temperature anisotropies in the CMB; nowadays these aspects of the theory, closely linked as they are to observations \cite{kn:calabrese}, are regarded as the most important.

However, Inflation is still called upon to explain (or at least render less mysterious) some much more basic observations. Among these is the strange fact that, while spacetime has an observably non-trivial geometry, it seems that, at cosmological scales, three-dimensional space \emph{does not}\footnote{The evidence for Euclidean spatial geometry is interpreted as such in the context of General Relativity. In this work, we propose to modify GR (a little), so, strictly speaking, we should not assert that there is strong evidence for Euclidean spatial geometry in the modified theory. However, it is not plausible that such a modified theory would conspire to mislead us, by somehow changing the CMB angular power spectrum, the BAO, and so on, in just such a manner as to cancel the effects of an actual non-trivial spatial geometry.}. Much effort has been expended on searches for the curvature of three-dimensional space, to no avail thus far; it is important to note that this result has now been confirmed by many \emph{different} lines of investigation \cite{kn:sunny2,kn:sunny3}, with remarkable precision (see \cite{kn:efstath,kn:sunny1} for clear general discussions and \cite{kn:sunny4,kn:eleonora,kn:camphuis} for the latest data on this).

We are entitled to ask why our spacetime has this extraordinary structure. And it is indeed extraordinary: the only way for the spatial sections to have an effectively Euclidean\footnote{We use ``Euclidean'' to mean ``devoid of non-trivial geometry'', (usually) avoiding the word ``flat'' because it has misleading connotations: we mean that the curvature is zero, but \emph{also} that any other measure of geometric non-triviality in three-dimensional space, such as intrinsic \emph{torsion}, should vanish.} intrinsic geometry is if the spacetime curvature, evaluated on them, resides entirely in their \emph{extrinsic} geometry. But, for technical reasons to be briefly reviewed later (see the footnote in Section 3, below), this means that our spacetime is foliated by spatial slices in a very unusual way, one not possible for a generic spacetime. This calls for an explanation. (Note that this is not the traditional ``Flatness Problem'': we are referring to the highly non-generic manner in which spacetime is foliated at the \emph{present} time, regardless of what this implies for the way in which the Universe has evolved.)

Inflation explains this strange structure by, first, supposing that the spatial curvature in the pre-Inflationary Universe was bounded; this is motivated by the standard hopes about the avoidance of singularities in some future complete theory of the earliest times. It is then plausible that the enormous, de Sitter-like expansion characteristic of Inflation reduced the region of space we can observe to such a small subspace (see \cite{kn:gibbons} for this) that it becomes indeed effectively Euclidean, apart from small perturbations due to quantum fluctuations. That sets up just the right initial conditions for the part of spacetime we can observe to evolve to its present state, with its unusual foliation.

Notice that this argument eliminates \emph{all} indications of non-trivial spatial geometry: it relies only on the assumption that the spatial sections have the structure of a differentiable manifold (and not, for example, that of a cone). That is, all \emph{classical} measures of non-Euclidean spatial geometry are ``inflated away'' (leaving only the effects of quantum fluctuations of the inflaton, the temperature anisotropies mentioned earlier).

If indeed this picture is correct ---$\,$ see below ---$\,$ it is often said that Inflation predicts that \emph{all} large-scale observations of the Universe should, on average, reveal structure which is \emph{isotropic} relative to some distinguished family of observers. One says that Inflation predicts\footnote{Penrose has argued \cite{kn:penrose} that, in fact, the existence of the Arrow of Time means that the earliest, pre-inflationary Universe was \emph{already} isotropic. While Penrose's argument is a strong one, and while nothing we will say contradicts that argument, it would be too confusing to pursue this here.} ``\emph{statistical isotropy}''. This specific \emph{prediction} of non-trivial spatial geometry being ``inflated away'' is the one with which we will be concerned here.

Now it should be noted that making this standard argument that ``Inflation implies isotropy'' reasonably rigorous is surprisingly difficult: see \cite{kn:pit} for a good discussion, particularly of the problems associated with the quantum-mechanical aspects. And indeed, various parts of this argument have, from time to time, been questioned. For example, it has sometimes been asked, in effect, whether standard Inflationary models really give rise to enough expansion to ``inflate away'' the geometry of three-dimensional space: see for example \cite{kn:guth} and particularly \cite{kn:pitrou}. From a completely different viewpoint, it has been pointed out that the field responsible for Inflation, the ``inflaton'', does not itself (necessarily) ``inflate away''; if, therefore, the inflaton is inherently anisotropic ---$\,$ if, for example, it is a vector field \cite{kn:soda1,kn:soda2} ---$\,$ then we might not expect Inflation to predict statistical isotropy. (See the discussion at the end of \cite{kn:pit}; see also \cite{kn:soda3} for some interesting recent developments in this field.)

This concept of ``vector Inflation'' builds anisotropy into Inflation from the outset, and arranges for that anisotropy to survive the extreme inflationary expansion. Here we will explain an alternative approach, in which the inflaton carries \emph{spin} \cite{kn:spinor}, but is \emph{not} inherently anisotropic \cite{kn:hehlquark}.

We shall therefore, henceforth, cleave to the standard picture, as described above: we will take it that Inflation does indeed ``inflate away'' all non-trivial \emph{intrinsic} spatial geometry, not approximately but (apart from inflaton-induced perturbations) essentially exactly, so that, at the end of Inflation, the spatial sections have an intrinsic geometry which is effectively exactly Euclidean, and therefore statistically isotropic. Our claim will be that, strange as it may seem, this picture can be reconciled with (some) observations of cosmic anisotropy.

All this has recently become topical, because statistical isotropy has in fact been cast into doubt by analyses of several data sets: see for example \cite{kn:eoin0,kn:eoin1,kn:eoin2,kn:eoin3,kn:ani,kn:eoin4,kn:wilt,kn:mig2,kn:maart,kn:boubel,kn:anis,kn:aniobserv,kn:salzano,kn:barjou,kn:eoinlow} (See also Section 4.6 of \cite{kn:cosmoverse} for a survey of these and other possible anomalies in the recent cosmic observations.) While this ``statistical anisotropy'' is by no means confirmed \cite{kn:scolnic}, it is difficult to believe that the agreement of all these diverse data is a ``fluke'' \cite{kn:eoin}. In any case, observational advances promise to settle the question in the not very distant future. Since these new observations challenge the very foundations of the standard understanding of the early Universe, they certainly deserve attention.

This prompts us to go back to basics and ask: is it really the case that (conventional) Inflation \emph{necessarily} predicts isotropic ``initial conditions'' (at reheating) for the subsequent evolution, leading to an isotropic Hubble field (the field representing the (directed) rate at which the galaxies separate as cosmic time passes)?

The key point here is that, in conventional models of Inflation, the intrinsic geometry of the spatial sections certainly \emph{is} ``isotropised'' \cite{kn:pit}: once we accept that all traces of non-trivial spatial geometry have been eliminated by Inflation, then the sections must be isotropic, simply because Euclidean space itself is indeed isotropic around any point. Thus reheating \cite{kn:reheat1,kn:reheat2,kn:reheat3}, which we can think of as the start of (relatively) ``well-understood physics'', does take place in an isotropic (Euclidean) three-dimensional space if conventional Inflation is valid.

Thus we are forced to re-formulate our question: does the isotropisation of the intrinsic geometry of the spatial sections at reheating imply an isotropisation of \emph{the full set} of ``initial conditions'', at that time?

This is not at all as obvious as it may look. The initial value problem for General Relativity requires a specification of both the intrinsic \emph{and} the \emph{extrinsic} geometry of the spatial sections. As we will review, the Hubble field is (primarily) a manifestation of the extrinsic geometry. The point is that it is far from clear that the extrinsic geometry of the cosmic spatial sections must necessarily behave in precisely the same way as their intrinsic geometry.

However, we have to contend with the fact that the intrinsic and extrinsic geometries are by no means entirely independent. To be more specific: the study of submanifolds in differential geometry requires us to specify two tensors on a spatial section $\Sigma$: the spatial metric $g$ and another bilinear form $h$. These two tensors are related in two ways.

First, they have to be related to each other and to the matter content of the spacetime by the initial value constraint equations \cite{kn:carlotto} if they are to generate a consistent temporal evolution from $\Sigma$. This evolution endows $h$, retrospectively as it were, with a geometric meaning, as the \emph{second fundamental form} of the embedding of $\Sigma$ into spacetime. Once this geometric meaning is established, it implies that there is a second, purely geometric relation between $g$ and $h$ (see Chapter VII of \cite{kn:kobnom2}).

In short, the intrinsic metric $g$ and the second fundamental form $h$ are related in various rather deep ways, both geometric and physical; so it is not out of the question that $h$ might \emph{inherit} isotropy from $g$, so that the full set of ``initial conditions'' bequeathed by Inflation is indeed necessarily isotropic.

Our objective here is to understand this precisely. We will find that, in all conventional Inflationary theories (meaning those in which anisotropy is not built in from the outset), and in all theories of gravitation using strictly metric spacetime geometry (meaning geometry which is completely fixed by a semi-Riemannian metric and nothing else), the extrinsic geometry \emph{does} necessarily share the symmetries of the intrinsic geometry: in all such theories, the Hubble field must isotropise if Inflation occurs. This is why it is so difficult to reconcile statistical anisotropy with (conventional) Inflation.

However, we also show that this is really a consequence of the special relationship between intrinsic and extrinsic geometries in strictly metric spacetimes, and that it is not the case in the simplest and geometrically most natural extensions of GR, those in which the \emph{torsion} is permitted to be non-zero \cite{kn:blaghehl,kn:lemos,kn:mavro}.

Our principal findings regarding cosmic isotropy in (what we shall call) ``\emph{Torsional Inflation}'' are as follows.

First, when torsion is present, it is indeed possible for a perfectly isotropic intrinsic spatial geometry to co-exist with anisotropy of the Hubble field immediately after Inflation: in other words, torsion supplies a natural way of reconciling Inflation with (\emph{some}) observations of statistical anisotropy.

Second, the inclusion of torsion gives rise to simple predictions as to the form taken by the Hubble field anisotropy: there is always just one distinguished direction, and that direction is always associated with a \emph{minimum} of the magnitude of the Hubble field. 

Third, torsion gives rise to a rotational component of the Hubble field. The extent of this rotation, which may however be difficult to detect, is correlated with the extent of anisotropy in the immediate post-inflationary Hubble field.

Fourth, the anisotropy we discuss here is mainly relevant to the end of the inflationary era, because it is due to the growth of the ``\emph{extrinsic torsion}'', which attains its maximum then. It follows that the anisotropy here is associated with perturbations corresponding to relatively \emph{small} cosmic scales. In other words, our work here does not explain very large-scale anisotropies: those are explained (still within the torsional Inflation picture) in a very different way, to be discussed elsewhere.

We begin with some simple observations regarding ordinary Inflation, in the context of standard General Relativity.

\addtocounter{section}{1}
\section* {\large{\textsf{2. What Inflation Does and Does Not }}}
To reiterate: the spatial sections of our Universe, those in which the galaxies are distributed approximately uniformly, appear (at present, and in the directly observable past) to be Euclidean, to a good approximation \cite{kn:efstath,kn:sunny1,kn:sunny4,kn:eleonora,kn:camphuis}. And yet, cosmic space\emph{time} is by no means close to being Minkowskian. Spacetime has a non-trivial geometry, but space does not. Some physical process has imposed this peculiar structure on spacetime. We assert that that process is what we call ``inflating away''.

In order to understand this ``inflating away'', let us begin by attempting to understand the two aspects, flatness and isotropy, separately. So we begin by assuming that no anisotropy is present\footnote{This may be more than a mere convenience: see the previous footnote regarding Penrose's argument \cite{kn:penrose} that the earliest Universe must have been already isotropic before Inflation.}, and focus at first on the fate of the spatial curvature. (In this Section, we use only standard General Relativity.)

The spacetime geometry during Inflation is then approximately that of de Sitter spacetime. Since we want the spatial curvature to be non-zero initially (so that we can study what happens to it), we choose the version of de Sitter spacetime which is foliated (globally) by spacelike three-spheres.

The scale factor in this case is $L\cosh(t/L),$ where $L$ is the length scale defined by the energy density (which does not vary) and $t$ is proper time, taken to satisfy $t\,\geq \,0$ ($t\,=\,0$ representing the beginning of Inflation.)

The \emph{intrinsic} spatial curvature tensor is\footnote{See \cite{kn:mci} for details of the following discussion; to ease comparison with the geometry literature, we mostly follow the notations of \cite{kn:kobnom2}.}
\begin{equation}\label{INTRINSIC}
R(W,\,Z,\,X,\,Y)\,=\, L^{-2}\m{sech}^2(t/L)\left(g(Y,\,Z)\,g(W,\,X)\;-\;g(X,\,Z)\,g(W,\,Y)\right),
\end{equation}
where $g$ is the spatial metric (that of a three-sphere here), $R$ denotes the spatial curvature tensor, and $W,\,Z,\,X,\,Y$ are spacelike tangent vectors. This evidently takes its maximum value at the start of Inflation, and then quickly decreases towards zero: the intrinsic curvature indeed ``inflates away'', exponentially quickly.

\emph{However}, the Hubble parameter $H$ does \emph{not} ``inflate away'': just the reverse:
\begin{equation}\label{HUB}
H \,=\,{1\over L}\,\tanh(t/L);
\end{equation}
we see that the Hubble parameter \emph{increases} during Inflation. This is our first indication that the Hubble field is not necessarily affected by Inflation in the same way as the intrinsic geometry of the spatial sections.

We can see the geometric significance of the $\tanh(t/L)$ factor by studying the \emph{extrinsic} curvature.

The Gauss formula for submanifold curvature (Section 4 of Chapter  VII of \cite{kn:kobnom2}) states that the full spacetime curvature, evaluated on tangent vectors to the spatial section $\Sigma,$ is simply the sum of the intrinsic and the extrinsic curvatures, evaluated on those same vectors. The total curvature is $L^{-2}\left(g^*(Y,\,Z)\,g^*(W,\,X)\;-\;g^*(X,\,Z)\,g^*(W,\,Y)\right)$, where $g^*$ is the spacetime metric (which can however be replaced here by the spatial metric $g$, since all of the vectors here are tangential to the spatial section); the coefficient is constant because the energy density does not change during Inflation. (This is of course the curvature tensor of de Sitter spacetime.)

Subtracting the intrinsic curvature, we see that the extrinsic curvature is given by
\begin{equation}\label{EXTRINSIC}
R^E(W,\,Z,\,X,\,Y)\,=\, L^{-2}\tanh^2(t/L)\left(g(Y,\,Z)\,g(W,\,X)\;-\;g(X,\,Z)\,g(W,\,Y)\right).
\end{equation}
Far from being ``inflated away'', the extrinsic curvature grows, in such a manner as to compensate for the decline of the intrinsic curvature. One might say that the extrinsic curvature \emph{has} to grow in order to make this compensation possible.

The coefficient, $L^{-2}\tanh^2(t/L)$, is of course the square of the Hubble parameter, which is thereby revealed to be \emph{an aspect of the extrinsic geometry} of the spatial sections. By the end of Inflation, the Hubble parameter has survived, indeed grown, because extrinsic geometry is immune to being ``inflated away''.

Notice that the extrinsic curvature is actually zero at $t\,=\,0$, the beginning of Inflation\footnote{One of the virtues of the ``no-boundary'' theory of the origin of the cosmos \cite{kn:lehners} is that it explains this initial geometry: the initial extrinsic curvature \emph{must} be zero in this theory.}. The situation at the start of Inflation was precisely the reverse of the one we find ourselves in today: at $t\,=\,0,$ the intrinsic curvature was large, while the extrinsic curvature was negligible, whereas currently the intrinsic curvature is (probably) too small to detect, while the extrinsic curvature is \emph{large}, manifested in the large value of the Hubble parameter at present.

In summary: what does Inflation do? It suppresses the \emph{intrinsic} geometry of the spatial sections, causing the spatial sections to have a Euclidean, isotropic structure.

What does Inflation \emph{not} do? It does \emph{not} suppress the \emph{extrinsic} geometry of the spatial sections: in fact, \emph{just the reverse}. The large Hubble field we see today parametrises that extrinsic geometry, and as such owes its existence to the fact that the extrinsic geometry ``inflates up'' instead of ``inflating away''.

In view of this, we should \emph{not} necessarily expect the extrinsic geometry to be isotropised by Inflation, and therefore we should not expect the Hubble field to be isotropised either. We can hope that, nevertheless, somehow it was: perhaps, in some way, the intrinsic geometry communicates its isotropisation to the Hubble field. But \emph{this is something to be proved}.

In order to proceed, we need a precise definition of the Hubble field, without assuming isotropy from the outset.

\addtocounter{section}{1}
\section* {\large{\textsf{3. Geometry of the Hubble Field }}}
We now give a mathematical formulation of the Hubble field. The exposition of the geometry here is adapted to our particular needs. For alternative expositions with substantially different emphases, see \cite{kn:brechet,kn:maart}.

We begin by supposing that spacetime has a semi-Riemannian metric $g^*$, and a connection $\nabla^*$ which is compatible with $g^*$ but which does not necessarily have zero torsion\footnote{We consistently use the asterisk superscript to denote spacetime quantities. The absence of the asterisk denotes a purely spatial quantity.}.

Throughout this work, we assume that spacetime can be foliated by spacelike hypersurfaces such that the galaxies are (statistically) uniformly distributed in \emph{all} of them, that is, for all time. Of course, this ``Copernican'' condition, often called ``homogeneity'', is a strong constraint.

Let $\Sigma$ be such a hypersurface and let $\xi$ be the corresponding local future-pointing timelike unit normal vector field. The integral curves of $\xi$ represent the distinguished observers, those who observe statistical homogeneity as above. We do \emph{not} assume that they observe an isotropic Universe.

We know that the distances between galaxies in our Universe are, overall, increasing functions of time: we refer to this fact as the \emph{cosmic expansion}. Let $X$ be a spacetime vector field which is tangent to each $\Sigma$, describing the spatial separation in each $\Sigma$ of two nearby objects, which we can take to be galaxies\footnote{To be more precise, let $p$ be a point in $\Sigma$ and let $s \rightarrow c(s)$ be a smooth curve in $\Sigma$ starting at $p$. Then $X$ is the tangent to this curve at $p$, and it represents the separation of nearby points along the curve from $p$. See the discussion of ``variations'' at the beginning of Chapter VIII of \cite{kn:kobnom2}.}. We call $X$ the \emph{separation field}. We need a way of measuring the rate of change of separation vectors, as seen by the distinguished observers. We will do this using differential-geometric language, so as to obtain a formulation which can easily be adapted to the case of spacetimes with torsion.

The galaxies can recede from each other in two distinct ways. First, they might actually be \emph{moving} relative to the distinguished observers (that is, changing their positions in $\Sigma$ ---$\,$ this is usually described in terms of ``\emph{peculiar velocities}''); and, second, they may not be moving, but the rules for computing spatial separations might themselves be time-dependent. This second kind of change of separations, the kind which is due \emph{exclusively} to geometric effects (that is, to the geometry of the way the foliation is embedded, to the geometry of the bulk spacetime, or to both ---$\,$ see below) is usually what we mean when we speak of ``expanding'' cosmological models, and it is this kind of time-dependent separation that concerns us here.

The obvious candidate for measuring the time rate of change of separations, the tangential component of $\nabla^*_{\xi}X$, is not satisfactory, because it includes the actual motion of the galaxies relative to each other. For example, in Minkowski space foliated by Euclidean spatial sections, $\nabla^*_{\xi}X$ is non-zero for any object moving relative to the corresponding observers, but the cosmological ``rate of expansion'' should be zero in this case. We need to find a way of removing those contributions to the rate of separation.

The way to do this is simply to subtract from $\nabla^*_{\xi}X$ the corresponding Lie derivative $\mathcal{L}_{\xi}X$ (before taking the tangential component). This works because the Lie derivative is not sensitive to the underlying geometry ---$\,$ it does not depend on the spacetime metric or connection: $\mathcal{L}_{\xi}X$ measures the non-geometric rate of change of $X$. This procedure will yield zero, as it should, in the above Minkowskian example; it amounts to stripping off any rate of separation due to actual \emph{motion} relative to the distinguished observers.

This idea is inspired by a construction well known in differential geometry: it is discussed (in more formal language) in Section 4 in Chapter VI of \cite{kn:kobnom1}, where the deep geometric meaning of this procedure is developed. (In particular, the object so defined (denoted by $A$ there) is used to map Killing vectors into the Lie algebra of the linear holonomy group of a Riemannian manifold; this is connected with Kostant's Theorem, Theorem 4.5 in Chapter VI of \cite{kn:kobnom1}.)

To conclude: the time evolution of galactic separations is measured by a mapping which we denote by $\hat{\frak{H}},$ defined by
\begin{equation}\label{A}
\hat{\frak{H}}(X)\,=\,\tau \left(\nabla^*_{\xi}X\,-\,\mathcal{L}_{\xi}X\right),
\end{equation}
where $\tau$ denotes the tangential component.

Note that, if $F$ is a smooth function on spacetime, then $\hat{\frak{H}}$ annihilates $F$, so in fact $\hat{\frak{H}}$ acts \emph{linearly} on linear combinations of scalar field multiples of spacetime vector fields (though neither the covariant nor the Lie derivative does so); in particular, it acts linearly with respect to vector and scalar fields on $\Sigma$, when $\hat{\frak{H}}$ is restricted to $\Sigma$. Thus $\hat{\frak{H}}$ is a \emph{linear} transformation which maps the separation vector to that part of the rate of change of the separation of the galaxies which is due to purely geometric effects.

In summary, the fundamental geometric rate of separation is related to the separation vector by a $(1,\,1)$ tensor\footnote{Here, and henceforth, we indicate a $(1,\,1)$ tensor by a ``hat''; the metrically equivalent $(0,\,2)$ tensor is denoted by the same symbol without the ``hat''.} on $\Sigma$. This we can call the \emph{generalised Hubble-Lema\^{i}tre Law}: as we will see in detail later, it reduces to the familiar Law in the cases where $\hat{\frak{H}}$ (which we will call the ``\emph{Hubble tensor}'') reduces to a simple multiple of the identity map. This is how we propose to give a geometric, coordinate-independent formulation of the Hubble-Lema\^{i}tre Law.

We stress that, by deliberate construction, the Hubble tensor depends on the cosmological spacetime and spatial \emph{geometry}. In particular, it is defined in terms of $\nabla^*,$ which in turn depends on the metric \emph{and the torsion} if the latter is not zero. We must therefore expect to find that, in the most general case, the cosmological expansion is not just a matter of having a spatial metric that depends on cosmic time. Instead, we should anticipate that the cosmic expansion is \emph{also} partially due to the direct geometric effects of torsion; and so it will prove. Thus, torsion can give rise to a kind of cosmic expansion which is radically different to the kinds with which we are familiar ---$\,$ including the kind that is central to Inflation. We will be more explicit about this below.

The $(0,\,2)$ version of $\hat{\frak{H}}$ is defined by writing
\begin{equation}\label{C}
\frak{H}(X,\,Y)\,=\,g(\hat{\frak{H}}(X),\,Y),
\end{equation}
where $X$ is as above, $g$ is the spatial metric (induced on $\Sigma$ by the spacetime metric $g^*$), and $Y$ is any tangent vector to $\Sigma$. Note that we are not assuming that $\frak{H}$ is a symmetric bilinear form, so the ordering here can be important.

Clearly $\frak{H}(X,\,n)$ represents, if $n$ is a unit spacelike vector, the component of the rate of separation of two galaxies with separation vector $X$, in the direction of $n$. In particular, then, $\frak{H}(X,\,X)$ is $|X|$ times the radial component of the rate of separation. Having discarded the peculiar velocities, we assume that this radial component is always positive for any $X \neq 0$. The motivation here is that we want to consider a Universe in which the distances to the galaxies never decrease or remain constant, in any direction. This does not mean, of course, that the expansion occurs \emph{at the same rate} in all directions or that the expansion is always radial. It simply reflects the fact that, peculiar velocities aside, cosmic \emph{contraction} is not observed.

Thus, the symmetric part of $\frak{H}$ (which is the part that determines $\frak{H}(X,\,X)$) is positive-definite, and it follows that the determinant of $\frak{H}$ itself is strictly positive: in particular it is a non-singular linear map.

We can understand $\frak{H}$ better by splitting it using the well-known \emph{polar decomposition} \cite{kn:hall}, which states that any non-singular bilinear form $B$ can be uniquely factorised as $B = Q\sqrt{B^TB},$ where $Q$ is an orthogonal linear map and where the square root denotes the positive square root of the positive-definite symmetric bilinear form $B^TB.$ This square root is itself a positive-definite symmetric bilinear form.

Applying this to $\frak{H}$, we therefore write
\begin{equation}\label{D}
\frak{H}\,=\,Q\,\eta,
\end{equation}
where $Q$ is an orthogonal transformation (in this case, since the determinant of $\frak{H}$ is positive, a rotation) depending on $\frak{H}$, and where\footnote{The letter $\eta$ was originally the Greek equivalent of $h$, though of course it no longer is.} $\eta$ is a symmetric positive-definite bilinear form or tensor on $\Sigma$, defined as the positive-definite symmetric square root of $\frak{H}^T\,\frak{H}$. Since, as we have seen, $\frak{H}$ is non-singular, $\eta$ is uniquely defined.

In the case of theories based on strict metric geometry, it turns out (see below) that $\frak{H}$ itself is a symmetric bilinear form. Because it is positive-definite, as discussed above, the uniqueness of the polar decomposition implies that $Q$ is the identity map in all such geometries.

In strictly metric geometry, then, the generalised Hubble-Lema\^{i}tre Law involves a simple symmetric bilinear form, $\frak{H}$. Its geometric interpretation is given by the spectral theorem: the eigenvectors of the corresponding linear transformation are mutually orthogonal, and the eigenvalues are positive real numbers, so the linear transformation represents a radial expansion by various amounts in three perpendicular directions. If the Hubble field is not isotropic, then we can quantify the anisotropy by studying the three real positive eigenvalues of $\frak{H},$ let us call them $H_i,\; i = 1,\,2,\,3,$ which will not all be the same; they describe the rates of separation in three perpendicular directions.

If the geometry is not strictly metric, then a rotational component, measured by $Q,$ can be present; from the above discussion it is clear that it \emph{must} be present if $\frak{H}$ is not a symmetric bilinear form, since $\eta$ is always symmetric. In three dimensions, a rotation always defines a distinguished direction, the axis of rotation. This axis need not coincide with any of the eigenvectors of $\eta.$ In general, then, the situation is rather complicated; but note that (since every real three-dimensional linear map has at least one real eigenvector with a real eigenvalue), in all cases there is at least one direction in which the rate of separation vector is radial.

Next, we need to explain the geometric meaning of $\frak{H}$ in more detail.

In the strictly metric case, this is straightforward. Since, in that case, we take the torsion tensor to vanish, we have, using the definition of that tensor (and the fact that $\mathcal{L}_{\xi}X = [\xi,\,X]$, the commutator of the two vector fields regarded as derivations),
\begin{equation}\label{B}
\hat{\frak{H}}(X)\,=\,\hat{h}(X),
\end{equation}
where $\hat{h}$ is defined by
\begin{equation}\label{DD}
\hat{h}(X)\,=\,\nabla^*_X\xi.
\end{equation}
Evidently $\hat{h}$ tells us how the unit normal $\xi$ changes as we move around in $\Sigma$. That is, it tells us the way in which space bends into spacetime. This is precisely what we call the \emph{extrinsic} geometry of the foliation: it is not at all concerned with the geometry inside the spatial slices, instead it is the study of the manner in which space ``fits into'' spacetime. We see, then, that the Hubble field is closely related to this extrinsic geometry: the cosmological expansion is (in the strictly metric case) due (exclusively) to \emph{the non-trivial bending of space into spacetime.}

In fact, $\hat{h}$ is the tensor on $\Sigma$ usually known \cite{kn:kobnom2} as the \emph{second fundamental form}. Notice that $\hat{h}(X),$ for any $X$ tangential to $\Sigma,$ is itself automatically tangential to $\Sigma$: one sees this by differentiating $g^*(\xi,\,\xi) = - 1$ and using the fact that the covariant derivative of $g^*$ is zero, so that $g^*(\xi,\,\nabla^*_X\xi) = 0,$ meaning that the normal component vanishes. (That is why we did not need to take the tangential component in equation (\ref{B}).)

As might have been expected, the second fundamental form controls the difference between the \emph{spacetime} curvature and the \emph{spatial} curvature: that difference is quantified by the \emph{extrinsic curvature tensor}, which is \emph{always}\footnote{We stress the ``always'': this holds true no matter how complex the spacetime geometry may be. This is in sharp contrast with the intrinsic curvature, which has such a simple form (as in equation (\ref{INTRINSIC})) only in very special cases. The Gauss formula for submanifold curvature means that, if the intrinsic curvature of the spatial sections vanishes, then the full spacetime curvature, when evaluated on $\Sigma,$ is forced to take this precise, very special, form. This is what we meant by saying earlier that the flatness of the spatial sections implies that the foliation of the Universe by spacelike hypersurfaces, as we observe it now, is exceedingly non-generic.} given by the formula
\begin{equation}\label{E}
R^E(W,\,Z,\,X,\,Y)\,=\, h(Y,\,Z)\,h(W,\,X)\;-\;h(X,\,Z)\,h(W,\,Y),
\end{equation}
for vectors $W,\,Z,\,X,\,Y$ tangential to $\Sigma$, and where here and henceforth $h$ is the $(0,\,2)$ version of $\hat{h}$. Note (see \cite{kn:mci}) that this expression for the extrinsic curvature (and also the Gauss formula relating it to the full spacetime curvature) is true for \emph{any} spacetime geometry endowed with a spacetime metric and a compatible linear connection: it is valid whether or not the geometry is what we are calling ``strictly metric''. (That is, it is valid even if $h$ is not a symmetric bilinear form; but in that case the ordering of the vectors in equation (\ref{E}) must be carefully preserved.)

Equation (\ref{B}) reveals that, in the strictly metric case at least, the Hubble tensor is just an aspect of the extrinsic geometry of the spatial sections. In principle, therefore, it is disconnected from the intrinsic geometry of the spatial slices; in particular, there is no reason at this point to think that the Inflationary isotropisation of the intrinsic geometry should ensure isotropy of the Hubble field, just as the ``inflating away'' of the intrinsic curvature does not imply any ``inflating away'' of the Hubble parameter ---$\,$ as we saw, this does \emph{not} happen.

Nevertheless, particular properties of strictly metric geometry \emph{do} in fact ensure that the Hubble field must share the symmetries of the spatial metric in that special case, as we now discuss.

\addtocounter{section}{1}
\section* {\large{\textsf{4. Isotropisation of the Hubble Field in Strictly Metric Inflation }}}
The description of a spatial hypersurface in spacetime involves two basic objects, the spatial metric $g$ and the second fundamental form $h$. In general, there is no particular reason to expect the two to be closely related; generically, for example, $h$ is not even a symmetric bilinear form \cite{kn:mci} at each point.

In the specific case of strictly metric geometry ---$\,$ recall that this means that the metric determines the geometry completely ---$\,$ however, there \emph{is} in fact a close relation: we have (see \cite{kn:mci}\cite{kn:kobnom2}),
\begin{equation}\label{F}
h\,=\,{1\over 2}\,\mathcal{L}_{\xi}\,g^*,
\end{equation}
where $\mathcal{L}_{\xi}\,g^*$ is the Lie derivative (evaluated on $\Sigma$) of the full spacetime metric $g^*$ with respect to $\xi$, the unit timelike normal to the spatial section, and it is understood that both sides are to be evaluated on a pair of vectors tangential to that section.

Although $g^*$ appears in equation (\ref{F}), actually the right side only expresses the ``time rate of change of the spatial geometry'', that is, of the spatial metric $g$. (One can see this most easily by expressing the restriction of $g^*$ to $\Sigma$ in the form $-\,\Xi \otimes \Xi \, +\, g,$ where $\Xi$ is the dual one-form to $\xi$.)

Notice that $h$ is a symmetric bilinear form in strictly metric geometry; and, since $\frak{H} = h$ here, so therefore is $\frak{H}.$ We see also, from equation (\ref{F}), that, \emph{in the strictly metric case}, the second fundamental form and the Hubble field are determined by the ``time rate of change of the spatial geometry''. In this case, then, it is valid to picture the expansion of the galaxies in terms of a ``stretching of space'', as of course we normally do.

The spatial section $\Sigma$ is said to be \emph{isotropic} if, given any pair of unit vectors at any point, there exists an isometry $\iota$ of $g$ (that is, a smooth map of $\Sigma$ to itself satisfying $\iota^* g = g$, where the asterisk denotes the pull-back), mapping one vector to the other.

Pick a point $p$ in a given spatial section $\Sigma$, and consider the integral curve $\gamma$ of $\xi$ passing through that point, so that $\gamma(0) = p.$ Let us suppose that $\Sigma$ is isotropic about $p$. (Of course, we do not insist on exact isotropy: all we ask is that the space near to $p$ is well-approximated by a three-dimensional space which is isotropic around $p$.)

Now, in Inflation, isotropy is approached asymptotically \cite{kn:pit}; that is, we are not interested in a situation where isotropy holds only at a single moment in time. Instead, we want the geometry around $\gamma(t)$ also to be approximately isotropic for at least some interval, of non-zero length, of parameter values along $\gamma$. This means that the Lie derivatives with respect to the Killing vectors associated with the isotropic mappings must commute with $\mathcal{L}_{\xi}$. Therefore equation (\ref{F}) implies that each isotropic mapping of $g$ is also an isotropic mapping of $h$, that is, of $\frak{H},$ since the two are equal in the strictly metric case.

In particular, if we assume that Inflation has isotropised $g$, then it also must isotropise $\frak{H}$. \emph{In strictly metric geometry, Inflationary isotropisation of the intrinsic geometry necessarily entails isotropisation of the Hubble field.}

Since $\frak{H}$ is a symmetric bilinear form here, and since the isotropy group acts irreducibly, Schur's lemma (see Appendix 5 of \cite{kn:kobnom1}) implies that, if the Hubble field is isotropic, then $\frak{H}$ must be a multiple (which, since we are assuming that the galaxies are always uniformly distributed, must not depend on position in $\Sigma$) of the spatial metric tensor $g$. That multiple is of course the Hubble parameter $H$, the thrice repeated positive eigenvalue of $\frak{H}:$
\begin{equation}\label{FF}
\frak{H}\,=\,H g.
\end{equation}
Since both $\frak{H}$ and $g$ are isotropic, so is $H$, in the obvious sense. Thus $H$ can at most be a function of time. Notice that this argument shows that the second fundamental form is also given by $h = H g.$

Of course, equation (\ref{FF}) means that $\hat{\frak{H}}$ acts simply by multiplication by $H$. Thus we see that our proposed definition of the Hubble field gives us a generalised Hubble-Lema\^{i}tre law which does indeed reduce to the classical law when we return to spatially isotropic General Relativistic cosmologies.

In the simplest cases one can easily confirm that $H$ is as expected. If we use Gaussian normal coordinates with a proper time coordinate $t$ (so that $\xi = \partial_t$) and set $g = a(t)^2\,f,$  where $a(t)$ is the scale factor and $f$ is a time-independent isotropic three-dimensional spatial metric tensor, then we have
\begin{equation}\label{G}
\frak{H}\,=\,{\dot{a}\over a}\,g,
\end{equation}
where the dot denotes differentiation with respect to proper time. Thus we regain the familiar expression $\dot{a}/a$ for the Hubble parameter in this case, and we see explicitly that $\frak{H}$ inherits isotropy from $g$.

In the case of the de Sitter spacetime foliated by spheres discussed above, we have
\begin{equation}\label{H}
\frak{H}\,=\,L^{-1}\tanh(t/L)\,g,
\end{equation}
where $g$ is the standard metric on the three-sphere, and again we see that $\frak{H}$ is isotropic because $g$ is so. Notice also that, as expected, during Inflation all measures of the extrinsic geometry, including the Hubble tensor/parameter, increase as the intrinsic geometry approaches Euclidean geometry.

Inflation in the case of a strictly metric spacetime, then, is a process whereby the Hubble tensor grows (from a very small or zero initial value) to a tensor which, by the time reheating begins, has a large magnitude and is as isotropic as the spatial metric.

There are two remarkable aspects of this argument. The first is that we have used the word ``Inflation'' in its broadest possible sense: as referring to an early phase of extremely rapid expansion that isotropises spatial geometry. The argument therefore applies to all of the many conventional (scalar) Inflationary models, and so the conclusion cannot be avoided by tinkering with the details (the precise nature of the inflaton potential or other features of its Lagrangian, and so on).

Second, and still more striking: the remainder of the argument is purely geometric ---$\,$ it makes no reference whatever to the field equations for gravity. It depends only on the assumption that the geometry of spacetime is strictly metric, an assumption that, without further input, leads inevitably to equation (\ref{F}). Thus, modifying the matter content or changing the Einstein-Hilbert action will make not the slightest difference to the argument, as long as we use strictly metric geometry.

\emph{We thus reach a very strong conclusion}: a confirmed observation of statistical anisotropy would mean either that

[a] All conventional models of Inflation ---$\,$ that is, all Inflationary models which induce isotropisation of the intrinsic spatial geometry ---$\,$ are seriously defective,

or that

[b] The geometry of spacetime is not strictly metric.

We wish to argue that pursuing option [a] should be attempted only if all else fails. If we accept this, then option [b] would be necessary if statistical anisotropy were to be confirmed observationally. The question is whether it is sufficient: does a non-strictly metric spacetime geometry permit a statistical anisotropy of some kind, even if Inflation has isotropised the intrinsic spatial geometry?

We will now argue that it does, and that in fact it does so in an extremely natural manner. Whether it can do so in a manner consistent with (all of) the observations is of course another matter. In fact, we will see that option [b] does permit some kinds of anisotropy, but not all.

\addtocounter{section}{1}
\section* {\large{\textsf{5. Torsional Inflation Naturally Generates Anisotropy }}}
The familiar assumption, in standard General Relativity, that the spacetime \emph{torsion} should vanish as a matter of principle, is almost impossible to justify\footnote{On the other hand, we feel that the assumption that the linear connection is a metric connection \cite{kn:lavinia} \emph{is} physically justified in most cases (the exception being the very special case in which the connection is a ``Schr\"{o}dinger connection'' \cite{kn:csil}, so that lengths of four-vectors are still globally well-defined, as they appear to be in our Universe ---$\,$ particle masses are just lengths of momentum four-vectors after all).  We always assume that our connections are metric connections. The reader who disagrees with this restriction can interpret it as a mere simplifying assumption; and it is certainly true that allowing non-metricity would complicate our discussion very substantially.} \cite{kn:blaghehl,kn:mci}. This is particularly clear in the case of cosmological spacetimes.

The idea that the linear connections of cosmological spacetimes might have non-zero torsion has a surprisingly long history: see \cite{kn:lehel1,kn:lehel2}. (See \cite{kn:boud,kn:bravo} for general discussions of torsional cosmology.) Current interest in spacetime torsion focuses on recent observational developments; for example, it has been suggested that torsion might be relevant to several of the cosmological ``tensions'' \cite{kn:tortens,kn:tong,kn:torcond}, particularly the notorious ``Hubble tension'' \cite{kn:realm,kn:sunny,kn:verde,kn:adam,kn:popolo,kn:onepercent}. Recently, the very interesting possibility that torsion might play an important role \emph{during Inflation} has also been explored; see for example \cite{kn:arko,kn:muzi}.

When it is evaluated on vectors tangential to $\Sigma$, the spacetime torsion for a given foliation by spacelike hypersurfaces $\Sigma$ splits, in a manner analogous to the familiar splitting of the curvature in Riemannian geometry, into an \emph{intrinsic} torsion (the torsion of the connection induced on $\Sigma$) plus an \emph{extrinsic} torsion \cite{kn:mci}, represented by a $(1,\,2)$ tensor given by
\begin{equation}\label{EXTORH}
T^E(X,\,Y)\,=\,\left(h(X,\,Y)\,-\,h(Y,\,X)\right)\xi \,=\, 2h^a(X,\,Y)\,\xi,
\end{equation}
where $h$ is, as before, the $(0,\,2)$ version of the second fundamental form (as given in equation (\ref{DD})), where $h^a$ is its antisymmetric part, where $X$ and $Y$ are tangential to $\Sigma$, and where $\xi$ is as before. The extrinsic torsion is the analogue of the extrinsic curvature, in the sense that both are determined by the manner in which $\Sigma$ is embedded: that is, both are determined by the second fundamental form. Note that $h$ is symmetric if and only if the extrinsic torsion is zero, and so we refer to $h^a$ as the \emph{extrinsic torsion two-form}.

The intrinsic and extrinsic torsions, together, control the translational component (see Appendix 2 of \cite{kn:mci}) of the holonomy transformation for closed loops confined to a given spatial section $\Sigma$: that is, they help to describe the consequences of parallel transport around purely spatial loops. They do not describe parallel transport around loops which begin in $\Sigma$ but then venture out beyond it into the ``bulk'' of the spacetime.

We characterised conventional Inflation as a process which explains, among many other things, the fact that we do not observe any evidence for non-Euclidean spatial geometry. In particular, Inflation converts intrinsic curvature to extrinsic curvature. In the same way, we expect that Inflation converts intrinsic torsion to extrinsic torsion, so that, by the end of the inflationary era (but not earlier), the intrinsic torsion should vanish.

In the preceding Section we saw that, in the absence of torsion, Inflationary spatial isotropisation necessarily entails an isotropic Hubble field. Now let us see what happens when torsion is allowed to be non-zero.

We immediately see that non-zero torsion affects our entire discussion in the previous section, because already we used the zero-torsion condition in moving from equation (\ref{A}) to equation (\ref{B}). Using the definition of spacetime torsion, $T^*(A,\,B) = \nabla^*_AB - \nabla^*_BA - [A,\,B]$ for arbitrary spacetime tangent vectors $A,\,B,$ we now have
\begin{equation}\label{I}
\hat{\frak{H}}(X)\,=\,\hat{h}(X)\,+\,\tau \, T^*(\xi,\,X),
\end{equation}
where $X$ is tangential to $\Sigma$, and where, as before, $\tau$ means that the component tangential to $\Sigma$ has been taken. (The same proof as before shows that $\hat{h}(X)$ is automatically tangential to $\Sigma$, but that is not necessarily true of $T^*(\xi,\,X),$ so we have to take the tangential component explicitly.)

Clearly the torsion can have a direct effect on the Hubble field. In a sense, this is only to be expected: the metric and the torsion play equal roles in determining the geometry of these spacetimes. However, this simple fact has the consequence that we can no longer (in general) picture the cosmic expansion as being due \emph{only} to the extrinsic geometry of the spatial sections (as controlled by $h$). Instead we have an additional term $T^*(\xi,\,X)$ which, as above, controls parallel transport around loops which leave $\Sigma$ and explore the ``bulk''.

Thus, we need to take into account the ``\emph{bulk torsion}'' $T^B$, a $(1,\,1)$ tensor given for our distinguished foliation by
\begin{equation}\label{MIX}
T^B(X)\, = \,T^*(\xi,\,X),
\end{equation}
where $X$ is tangential to $\Sigma$.

It will be convenient to decompose the bulk torsion into a part that is tangential to $\Sigma$, together with a part normal to it. We define a $(0,\,2)$ tensor $\beta$ on $\Sigma$ by
\begin{equation}\label{MIXX}
\beta(X,\,Y)\,=\,2\,g\left(\tau T^B(X),\,Y\right),
\end{equation}
the factor of 2 being included for later convenience. In terms of the $(0,\,2)$ versions, we have now from equation (\ref{I}),
\begin{equation}\label{MIXXX}
\frak{H}\,=\,h\,+\,{1\over 2}\beta.
\end{equation}
Note that $\beta$ has no particular symmetry property, as a bilinear form, in general.

The normal component of $T^B$ is measured by a ``bulk torsion one-form'' $b$ on $\Sigma$ defined by
\begin{equation}\label{MIXXXX}
b(X)\,=\,-\,2 g^*\left(\xi,\,T^B(X)\right);
\end{equation}
it does not affect the Hubble tensor, so it is of less interest than $\beta$.

Clearly $\frak{H}$ need not be a symmetric bilinear form here, but the relevant parts of our previous discussion continue to apply, since we did not assume symmetry there. Motivated by the absence of cosmic blueshifts, we continue to assume that the symmetric part of $\frak{H}$ is positive-definite, whence we conclude that the matrices of $\frak{H}$ are non-singular and therefore have a unique polar decomposition, allowing us to define a unique $\eta$ tensor, constructed from $\hat{\frak{H}},$ just as before: $\eta\,=\,\sqrt{\frak{H}^T \frak{H}}$.

The spacetime \emph{contortion tensor} $K^*$ is defined as the difference between the spacetime linear connection $\nabla^*$ and the fictitious connection $\nabla^0$ that spacetime would have had if the torsion had been zero:
\begin{equation}\label{CONTORTA}
K^*(A,\,B) \,\equiv \, \nabla^*_A B - \nabla^0_A B,
\end{equation}
for any spacetime vector fields $A,\,B$: it is the measure of the extent to which the geometry deviates from being strictly metric. Using the definition of the torsion tensor, one finds that
\begin{equation}\label{CONTORTB}
K^*(A,\,B) \,- \, K^*(B,\,A)\,=\, T^*(A,\,B).
\end{equation}
The contortion is not in general symmetric or antisymmetric in these ``slots''; on the other hand, since the covariant derivative of the metric tensor is zero with respect to both connections, the $(0,\,3)$ version of $K^*$ must be antisymmetric in two other slots:
\begin{equation}\label{CONTORTC}
g^*(C,\,K^*(A,\,B))\,=\,-\,g^*(B,\,K^*(A,\,C)).
\end{equation}
where $A,\,B,\,C$ are arbitrary spacetime vector fields.

Now by the definition of $K^*$, we have, for any vector field $X$ tangential to any of our distinguished sections $\Sigma,$ $\nabla^*_X\xi \,=\, \nabla^0_X\xi \,+\,K^*(X,\xi)$; and so, from the definition of the second fundamental form, and making use of equation (\ref{CONTORTB}), we have
\begin{equation}\label{K}
h(X,\,Y)\,=\,{1\over 2}\,\mathcal{L}_{\xi}\,g^*(X,\,Y)\,+\,g\left(\tau K^*(\xi,\,X),\,Y\right)\,-\,g\left(\tau T^*(\xi,\,X),\,Y\right),
\end{equation}
for arbitrary vector fields $X,\,Y$ tangential to $\Sigma$. This is the torsional generalisation of equation (\ref{F}).

From equation (\ref{CONTORTC}) it follows that the second term in equation (\ref{K}) defines a two-form on $\Sigma:$
\begin{equation}\label{P}
\varkappa(X,\,Y)\,=\,2\,g\left(\tau K^*(\xi,\,X),\,Y\right).
\end{equation}
We will call this the \emph{contortion form}.

Using the definition of $\beta$ (equation (\ref{MIXX})) we can write equation (\ref{K}) in the form
\begin{equation}\label{PP}
h\,=\,{1\over 2}\,\mathcal{L}_{\xi}\,g^*\,+\,{1\over 2}\,\varkappa \,-\, {1\over 2}\beta.
\end{equation}

Equation (\ref{PP}) is the fundamental geometric relation here. It tells us that the embedding of a spacelike hypersurface into spacetime is determined by the time rate of change of the intrinsic spatial metric, \emph{together} with two additional terms describing the effects of torsion.

Combining equations (\ref{MIXXX}) and (\ref{PP}), we find
\begin{equation}\label{M}
\frak{H}\,=\,{1\over 2}\,\mathcal{L}_{\xi}\,g^*\,+\,{1\over 2}\varkappa\,;
\end{equation}
here, as ever, it is understood that all of these objects are to be evaluated on pairs of vectors tangential to $\Sigma$.

Thus, $\varkappa$ is the part of the torsion that directly influences the Hubble field; it is just (twice) the antisymmetric part of the Hubble tensor.

Now let us focus on the Inflationary era, and let us follow the discussion of the previous section as closely as possible. The guiding principle here is that, after all, the absence of torsion from that discussion was incidental (or, better, an historical accident \cite{kn:blaghehl}).

We agreed earlier to accept that Inflation erases all forms of non-trivial intrinsic geometry in $\Sigma$, including the intrinsic torsion (the torsion of the induced linear connection on $\Sigma$). By the same logic, we take it that the measures of the extrinsic geometry of the spatial sections have \emph{increased}.

More specifically, we mentioned earlier that, in the ``no-boundary'' theory \cite{kn:lehners} of the earliest Universe, the extrinsic curvature is zero at the ``beginning of time''; and this is indeed a natural condition to impose at the start of Inflation, whether or not one accepts the ``no-boundary'' theory. For otherwise we would have to explain why the extrinsic geometry takes some specific form at the beginning, and this means importing a large number of seemingly arbitrary parameters into the theory.

The same reasoning applies here, and so, if we accept this line of argument, then \emph{both} the extrinsic curvature \emph{and} the extrinsic torsion vanish at the beginning. Both must however grow as Inflation proceeds, so, in particular, the extrinsic torsion $T^E$ will not be negligible by the time of reheating.

As before, the spatial metric $g$ will, with these assumptions, be isotropised by Inflation. Then the first term on the right of equation (\ref{PP}) isotropises, by the same argument as in the strictly metric case, and so it takes the familiar\footnote{Note however that $H$ no longer has, in general, a direct interpretation in terms of the eigenvalues of the Hubble tensor.} form $Hg$. But does this mean that $h$ and $\frak{H}$ also isotropise?

If $h$ isotropises, then so does its antisymmetric part, $h^a$. Now it can be shown (Appendix 5 of \cite{kn:kobnom1}) that a real antisymmetric bilinear form can be isotropic, in odd numbers of dimensions (3 in our case here), only if it is zero.

However, from equation (\ref{EXTORH}), we see that this would mean that the extrinsic torsion has to vanish. So we conclude that \emph{the extrinsic geometry of the distinguished spatial sections cannot isotropise} as a result of Torsional Inflation ---$\,$ just the reverse.

\emph{This is the central observation of the present work.} Our hypothesis is that, just as standard Inflation is essentially a process which transforms intrinsic to extrinsic curvature, so also does it convert intrinsic to extrinsic torsion, and so, as we just saw, the extrinsic geometry at the end of Inflation cannot be isotropic. It follows that the Hubble field at the time of reheating \emph{cannot} be isotropic in Torsional Inflation. This effectively means that the ``initial'' conditions for the post-inflationary evolution are inherently anisotropic, and of course this affects the later geometry. We are proposing this as a natural explanation of the purported observations of anisotropies beyond those due to quantum fluctuations. Since the extrinsic torsion is maximal towards the end of Inflation, the relevant anisotropies pertain to relatively small scale anisotropies, since those are the ones that are related to inflationary fluctuations at that time.

If we \emph{force} $h$ and $\frak{H}$ to be isotropic, as in \cite{kn:barrow}, then the extrinsic torsion vanishes identically throughout Inflation, which of course is not what we want here. If nevertheless we persist in enforcing isotropy, then, from equation (\ref{M}), we see that $\varkappa = 0,$ and so $\frak{H}\,=\,Hg$ as in ordinary General Relativistic cosmology. Since $h$ now has to be symmetric, equation (\ref{PP}) now forces $\beta$ to be likewise symmetric, and we have $h\,=\,{1\over 2}\,\mathcal{L}_{\xi}\,g^*\,-\, {1\over 2}\beta.$ Schur's lemma implies that $\beta$ is a possibly time-dependent multiple, let us (following \cite{kn:barrow}) call it $- 4\phi(t)$, of $g$.  In this case, then, we have
\begin{equation}\label{BARROW}
h\,=\,\left(H\,+\,2\phi\right)g.
\end{equation}
Notice that, in isotropic torsional theories, the second fundamental form differs from the Hubble tensor: this is due to the fact that the bulk torsion, unlike the extrinsic torsion, is not forced to be zero in the fully isotropic case.

Note that the bulk torsion one-form $b$, given by equation (\ref{MIXXXX}), must vanish in the isotropic case because it is metrically dual to a purely spatial vector field. However, spatial isotropy does \emph{not} force the intrinsic torsion to vanish: it might be a constant multiple of the spatial volume form (which means that it must be completely antisymmetric in all pairs of ``slots''). In \cite{kn:barrow} it is assumed from the outset that the intrinsic torsion is zero; presumably this is for simplicity. Here we do not assume this: on the contrary, we assume that the intrinsic torsion was initially \emph{large} (in the sense of being a large multiple of the volume form), and only approaches zero towards the end of the inflationary era.

The special case of ``isotropic torsion'' is the case usually assumed in discussions of torsional cosmology: see \cite{kn:barrow}. As we have seen, isotropy in torsional cosmology means that the only surviving components of the spacetime torsion are $\beta$ (which must be symmetric) and the intrinsic three-dimensional torsion $T$ (which must be completely antisymmetric). The Hubble law is exactly as in General Relativity, but the second fundamental form is sensitive to the presence of $\beta$ (equation (\ref{BARROW})): this means that the torsional version of the Friedmann equation, obtained by inserting the right side of equation (\ref{BARROW}) into the relevant constraint equation \cite{kn:luz} does differ from the General Relativistic Friedmann equation. (The result is precisely the generalised Friedmann equation proposed in \cite{kn:barrow}.)

The question as to what happens if we \emph{force} torsion to share the symmetries of the metric has been studied in a more general and complete manner, leading to important classification theorems: see \cite{kn:krss,kn:coley0,kn:coley1,kn:coley2}. Our point here, however, is just that \emph{Inflation alone does not impose such high degrees of symmetry on spacetime torsion}.

In summary, then, we have, if the intrinsic spatial geometry is isotropised, then the Hubble tensor takes the form
\begin{equation}\label{JUDE}
\frak{H}\,=\,H g\,+\,{1\over 2}\varkappa\,,
\end{equation}
while the second fundamental form is
\begin{equation}\label{JUDEE}
h\,=\,H g \,+\,{1\over 2}\,\varkappa \,-\, {1\over 2}\beta.
\end{equation}
Notice the way that the Hubble parameter $H$ \emph{mixes with the various components of the torsion}.

All this applies to the most general torsional geometries (for a connection compatible with a spacetime metric). Now we need to choose the specific kind of torsional geometry that is best adapted to torsional Inflation.

We saw earlier that, from a geometric point of view, standard Inflation (in the zero-torsion case) is just a mechanism for converting intrinsic curvature to extrinsic curvature. We take the view that precisely the same should be true in the torsional case, and so, henceforth, we will focus on the case where the intrinsic and extrinsic torsions are the \emph{only} non-zero torsion components. That is, \emph{we will only consider the case in which the bulk torsion is always absent.}\footnote{Note that this assumption is also made in the many works (see for example \cite{kn:lasenby} in which the spin density is associated with a \emph{Weyssenhoff fluid}. However, in that case, the intrinsic torsion is also assumed to vanish at all times, and of course that would be too strong here.}

This means that we will henceforth impose the conditions (see equations (\ref{MIXX}) and (\ref{MIXXXX}))
\begin{equation}\label{TORINFLAT}
\beta\,\equiv \, 0,\;\;\;b\,\equiv \, 0.
\end{equation}
Comparing equations (\ref{JUDE}) and (\ref{JUDEE}) we see that this means that, henceforth, we always have $\frak{H} \,=\, h.$ That is, the Hubble field is identified with the second fundamental form: the expansion is purely due to the manner in which the distinguished spatial sections are embedded, precisely as is the case in standard General Relativity. In this sense, the cosmologies we are considering are the simplest torsional extensions of the standard FRW cosmologies.

Combining equation (\ref{EXTORH}) with the antisymmetric part of equation (\ref{PP}), one has
\begin{equation}\label{CATE}
\varkappa \,=\,\beta^a \,+\,2 h^a,
\end{equation}
where $\beta^a$ is the antisymmetric part of $\beta,$ where $h^a$ is the extrinsic torsion form discussed earlier, and where it is to be understood that all three terms are evaluated on pairs of vector fields tangential to $\Sigma.$

If we enforce equation (\ref{TORINFLAT}), then, from equation (\ref{CATE}), we see that $\varkappa$ is $2 h^a\,$. We therefore encourage the reader to think of $\varkappa$ as being just (twice) the extrinsic torsion.

Clearly it is natural to propose that torsion may help us to reconcile Inflation with purported statistical anisotropies of various kinds: torsion provides us with a large (though, as we will see, far from limitless) repertoire of techniques to induce anisotropy in those parts of the Hubble field not controlled by the intrinsic geometry of spatial sections.

When we assume that equation (\ref{TORINFLAT}) holds, equation (\ref{JUDEE}) shows that the Hubble parameter is one third of the trace of the second fundamental form, which is usually called the mean curvature\footnote{By an odd coincidence, the mean curvature is usually denoted $H$ in the differential geometry literature.}. This might be said to supply a geometric meaning for $H$.

We have now a rather complete understanding of the consequences of the fact that the extrinsic geometry of a torsional spacetime need not be (and in fact, under reasonable assumptions, cannot be) isotropic. We now ask: what kinds of anisotropy are possible in this picture?

\addtocounter{section}{1}
\section* {\large{\textsf{6. Torsional Cosmological Anisotropies}}}
It is important to understand that \emph{not every form of cosmic anisotropy can be realised by torsion}. For example, the kind of theoretical anisotropy we discussed in the purely metric case (which is possible in the absence of Inflation), where there is radial separation (at three different rates) along three mutually perpendicular axes, is \emph{impossible} in Torsional Inflation with $\varkappa \neq 0$. This follows from the simple fact that a real $3 \times 3$ matrix with three mutually perpendicular eigenvectors is necessarily symmetric, and we know that $\frak{H}$ is not a symmetric bilinear form under our stipulations.

Let us be more explicit regarding the precise form that anisotropy takes in Torsional Inflation.

Since $\varkappa$ is antisymmetric, we have, from equation (\ref{JUDE}),
\begin{equation}\label{U}
\frak{H}^T \frak{H} \,=\,H^2g\,-\,{1\over 4}\varkappa^2.
\end{equation}
The matrix of $\varkappa$ is a real antisymmetric $3 \times 3$ matrix, so, if it is not zero, it has one eigenvector with zero eigenvalue, and two other eigenvalues which are pure imaginary and mutually conjugate. It follows that there exists a real orthonormal basis, with the zero-eigenvalue eigenvector as the third member, with respect to which the matrix of $\varkappa$ is
\begin{equation}\label{V}
[\varkappa]\,=\, \left(\begin{array}{ccc}
  0 & \kappa &0  \\

 - \kappa & 0 & 0 \\
 0 & 0 & 0 \\
  \end{array}\right),
\end{equation}
where $\kappa$ is a real quantity, the \emph{contortion parameter}, which, since we are assuming spatial homogeneity at all times, is constant in space though not necessarily in time. In this basis, $\varkappa^2$ has a diagonal matrix with entries $- \kappa^2,\,- \kappa^2,\,0,$ and so the matrix of the $\eta$ tensor (the square root of $\frak{H}^T \frak{H}$) is, from equation (\ref{U}),
\begin{equation}\label{W}
[\eta]\,=\, \left(\begin{array}{ccc}
  \sqrt{H^2 + \kappa^2/4} & 0 &0  \\

 0 & \sqrt{H^2 + \kappa^2/4} & 0 \\
 0 & 0 & H \\
  \end{array}\right).
\end{equation}
We see that $\varkappa$ defines a distinguished direction, the direction of its eigenvector with zero eigenvalue; this is also the direction in which $\eta$ has an eigenvector with eigenvalue $H$. Furthermore, it is not difficult to show that the rotation transformation, $Q,$ in this case has a matrix, relative to this basis, of the form
\begin{equation}\label{WWW}
[Q]\,=\, \left(\begin{array}{ccc}
  {H\over \sqrt{H^2 + \kappa^2/4}} &  {- \kappa/2\over \sqrt{H^2 + \kappa^2/4}}  &0  \\

 {\kappa/2\over \sqrt{H^2 + \kappa^2/4}} & {H\over \sqrt{H^2 + \kappa^2/4}} & 0 \\
 0 & 0 & 1 \\
  \end{array}\right).
\end{equation}
As we foresaw, this rotation is \emph{always} non-trivial if the contortion does not vanish.

The rotation is clearly around an axis which \emph{coincides} with the distinguished axis of the $\eta$ tensor. This distinguished direction is the unique real eigenvector of the Hubble tensor, the eigenvalue being of course $H$. We stress that the rate of separation vector is never radial except in the distinguished direction, even though the spatial geometry is isotropic.

Whether a non-radial ``expansion'' at the end of Inflation could leave a detectable signal in the physics at that time is not clear. (For discussions of the physics of non-radial astronomical motions in a different context, see \cite{kn:peacock,kn:transverse}.) The point however is that such an effect \emph{must} be present if the extrinsic torsion is, as we expect, non-zero at that time.

In more detail: let us define a ``Hubble eccentricity'' $\m{e}$ by
\begin{equation}\label{WW}
\m{e} \,=\, {\kappa \over H};
\end{equation}
This gives us a dimensionless measure of the anisotropy; it is also a measure of the relative importance of contortion (that is, under our assumptions, of extrinsic torsion) compared to extrinsic curvature.

In terms of $\m{e},$ the angle of the ``Hubble twist'' is given by
\begin{equation}\label{PSII}
\psi\,=\,\arccos\left({1\over \sqrt{1 + \m{e}^2/4}}\right).
\end{equation}
Thus indeed $\kappa$, in the form of the Hubble eccentricity, determines the magnitudes of both effects.

To conclude this discussion\footnote{Bear in mind throughout this discussion that we are considering the state of affairs at the end of Inflation, not in the later Universe. Thus for example ``Hubble parameter'' refers to its value at that time, not at present.}: we can see four facts very plainly, regarding Torsional Inflation (in the generic case, with $\varkappa \neq 0$).

[a] The predicted anisotropy singles out a distinguished direction; the Hubble parameter $H$ is the rate of separation in that direction.

[b] We saw earlier that not every form of anisotropy can be explained by contortion or extrinsic torsion; part of the reason was that the anisotropy always (in Torsional Inflation) has a rotational component. A key point here is that the parameter $\m{e}$ that governs the magnitude of the anisotropy is the \emph{same} parameter that determines the Hubble twist. Given one, we can predict the other unequivocally, and in particular it is not possible to have one without the other.

[c] The magnitude of the Hubble field during Inflation \emph{always} has its \emph{minimum} in the distinguished direction. The maxima occur in all directions at right angles to the distinguished direction, and the ratio of the maximum magnitude of the rate of separation vector to its minimum is $\sqrt{1 + {\m{e}^2\over 4}}.$

[d] The Hubble field during Inflation is isotropic in the planes perpendicular to the distinguished direction: that is, the isotropy group has been ``broken'' by torsion, but part of it still survives. The rotation $Q$ is in this case just a rotation of those isotropic planes.

[e] Since the extrinsic torsion grows throughout Inflation, these anisotropies in the Hubble field will be important (if at all) only towards the very end of the inflationary era. If they affect the inflationary perturbations, the effect will be seen on those modes which leave the horizon last: that is, on relatively small scales. Thus it seems that this aspect of torsional Inflation pertains to anisotropies at those scales, not on the very largest scales.

\addtocounter{section}{1}
\section* {\large{\textsf{7. Some General Observations on Torsion and Anisotropy}}}
It is well known that, except in special cases, the worldlines of photons and free massive particles (whether endowed with spin or not) in torsional theories are not ``straight'', that is, not geodesics, as they are in General Relativity. This seems unnatural at first sight. We wish to argue that \emph{the close relation between torsion and anisotropy discussed above removes this problem}.

We begin with a discussion of the equation of a photon worldline in torsional theories. (The analysis in this digression is based on the relevant discussions in \cite{kn:santana,kn:boud}.)

Let $u$ be the tangent vector, usually called the four-dimensional wave vector, to a null curve with a parameter $\lambda$ (which we can modify if necessary) in a spacetime with torsion (of any kind). In geometric optics, $u$ can be expressed as the four-dimensional gradient of a scalar eikonal function $\varphi$; that is, the one-form dual to $u$ (using the spacetime metric $g^*$), let us call it $U,$ can be written as $U = d\varphi.$ This is true not just for the standard Maxwell equations, but also for a very large family of generalisations of those equations \cite{kn:santana}.

As before, we denote by $\nabla^0$ the fictitious connection defined purely by the spacetime metric $g^*$: it is the connection that spacetime would have had if the torsion had been zero. The Hessian of $\varphi$ relative to $\nabla^0$ is $\nabla^0d\varphi = \nabla^0U$, and this is a symmetric bilinear form (because the torsion of $\nabla^0$ is zero):
\begin{equation}\label{OMICRON}
\nabla^0_AU (B)\,-\,\nabla^0_BU (A) \,=\, 0,
\end{equation}
where $A$ and $B$ are arbitrary spacetime vectors. Note that the validity of this equation relies on the fact that $U$ is an exact form; that is, on the assumption that $u$ is expressible in terms of an eikonal.

Now set $A = u;$ then using the fact that the metric has zero covariant derivative with respect to $\nabla^0$ we find that the second term on the left becomes $- \,u \cdot \nabla^0_Bu,$ and this is zero because $u \cdot u$ is a constant (zero in this case). (Here, for notational clarity, we are using the dot to represent the semi-Riemannian scalar product defined by $g^*.$) Therefore $\nabla^0_AU (B) = 0$ for arbitrary $B$, so $\nabla^0_uU\,=\,0,$ which means
\begin{equation}\label{PI}
\nabla^0_uu = 0.
\end{equation}
This is the equation for the worldlines of photons.

It is also the equation for the worldline (parametrised by the spacetime interval in this case) of a free massive particle. To see this, recall that, in relativistic Hamilton-Jacobi theory \cite{kn:synge}, the four-momentum of such a particle can be expressed locally as the gradient of the relativistic Hamilton principal function. The argument then proceeds as above, since $u \cdot u$ is a constant in this case also.

Equation (\ref{PI}) has of course a well-known geometric meaning: it means that massive free particles have worldlines which extremise the proper time between given spacetime events (and a suitably modified statement applies to photons). In general, however, in the torsional case, equation (\ref{PI}) also means that these worldlines are \emph{not} spacetime geodesics. That is, the worldlines of such particles are \emph{not always straight}: instead we have, from the definition of the spacetime contortion $K^*$,
\begin{equation}\label{PIA}
\nabla^*_uu\,=\,K^*(u,\,u),
\end{equation}
and the right side is not always zero; so the worldlines of photons and of massive free particles are not \emph{always} geodesics when the contortion is non-zero.

All this is well known, but we have derived it here in order to emphasise that it can indeed be \emph{derived} purely from the presumably uncontroversial assumptions that $u$ has a fixed length and is given by the gradient of an eikonal or Hamilton principal function. We do not need to assume that we are using any specific torsional theory, such as the Einstein-Cartan theory; we do not even need to ask whether the spin of photons or free massive particles ``couples'' to torsion, and so on.

Nevertheless it seems at first rather strange that torsional theories can apparently violate something as basic as Newton's First Law (``A free particle moves along a straight line at constant speed.'') ---$\,$ which is essentially what we are saying here. But this is where the study of anisotropy comes to the fore. The intuition that free particles should have straight worldlines is built on the assumption that space and time are \emph{uniform}: a free particle cannot in that case ``justify'' deviating from straight line motion, or moving at a non-constant speed. This assumption of uniformity seemed natural to Newton, and it also holds true in Special Relativity, where spacetime is maximally symmetric; but, even in ordinary General Relativity, it does not. For we know that, in that theory, three-dimensional space is \emph{not} ``uniform'' in general ---$\,$ spacelike Killing vectors do not always exist in all directions, and this can play a fundamental role in the physics, as it does for example in the Schwarzschild geometry. Similarly, time is not ``uniform'' in spacetimes, such as cosmological spacetimes, that lack a timelike Killing vector field.

Thus, this version of the First Law rests on very doubtful foundations even in ordinary General Relativity. In torsional theories, where, as we have seen, some kind of spatial non-uniformity is almost inescapable, we have even less reason to think that worldlines of free particles should be straight, so there is really no occasion for surprise when we find that they (often) are not\footnote{Newton's First Law can (and should) be salvaged by reformulating it: by declaring that a particle on which no external influence acts should have an action defined in terms of quantities pertaining only to the particle itself (its mass if any, the length or canonical parameter along its own worldline). The principle of stationary action then leads to the correct equation for the worldline, that is, to equation (\ref{PI}), \emph{provided} that we accept that gravitation is \emph{not} an ``external influence'' (that is, there is no additional term in the particle Lagrangian representing gravity). In this way we can preserve the greatest conceptual achievement of General Relativity, the abolition of gravitational ``force''.}.

We can see this explicitly by considering the special case of ``isotropic torsion'', discussed earlier. We saw that $\beta$ and the intrinsic torsion $T$ are the only surviving components of the torsion in that case; that $\beta$ is given by $\beta = - 4\phi(t)g$ (where $\phi(t)$ is some function of time only) and that $T$ had to be totally antisymmetric. When these specific forms are inserted into the right side of equation (\ref{PIA}), one finds that $T$ drops out because of the symmetry in $u$, while $\beta$ gives rise to a term (along the worldline) of the form\footnote{In the case of a photon, the right side takes the form $2\phi\omega u,$ where $\omega$ is the frequency. Here $\phi$ and $\omega$ are evaluated along the worldline, so they can be regarded as functions of $\lambda$.} (function of the parameter $\lambda$) $\times\, u$.

In the case of a massive particle, this means that the right side of equation (\ref{PIA}) must vanish, because the spacetime interval is always a canonical parameter in the timelike case. In the case of a photon, it means that the right side can be reduced to zero by a suitable change of parameter. We conclude that the worldline of a photon or of a free massive particle in the isotropic case is in fact a geodesic.

Conversely, then, if the worldline of a free particle is not a geodesic, then that is to be explained by some kind of anisotropy, due to the presence of torsion. This is the principal point we wish to make here: \emph{the failure of free particles and photons to have geodesic worldlines in torsional theories is simply due to the natural association between torsion and anisotropy}.

As a final aside, it is natural to ask whether the observers who regard the spatial sections as homogeneous (that is, those whose tangent vectors are given by $\xi$) are free particles. This is not altogether obvious; think, for example, of the ``Schwarzschild observers'' outside a Schwarzschild black hole, who do have to exert a force in order to appreciate the fact that the spacetime has a timelike Killing vector field. Thus it would not be very surprising if it turns out that our distinguished observers need to exert themselves in order to perceive homogeneity.

However, it is known (see \cite{kn:henk,kn:gour}) that homogeneity \emph{alone} (that is, without any assumption of isotropy) suffices to ensure that the relevant \emph{lapse function} is constant on each spatial section, and this in turn means that $\xi$ has to satisfy $\nabla^0_{\xi}\xi = 0.$
That is, the worldlines of the ``homogeneous observers'' do correspond to free particles. Again, their paths through spacetime are not usually straight, but now we understand that this is due to the anisotropy of the extrinsic geometry associated with torsion.

\addtocounter{section}{1}
\section* {\large{\textsf{9. Conclusion }}}
We live in a time of cosmological uncertainty \cite{kn:cosmoverse}. There are several ways to respond to this; among them, to search for alternatives to General Relativity.

These alternatives are often themselves classified according to the degree to which they are ``radical''. We have argued \cite{kn:mci} that the \emph{least} radical alternative to General Relativity is to consider non-zero torsion$\,$: because this is hardly an ``alternative'' at all.

To see this very briefly, we remind the reader that the equation of geodesic deviation for an arbitrary connection $\nabla$ takes the form (see \cite{kn:kobnom2}, Chapter VIII)
\begin{equation}\label{PSI}
\nabla_u \nabla_u D\; +\;\,\nabla_u\left[T(D, u)\right] \;+\;R(D, u)u \;=\;0,
\end{equation}
where $u$ is the tangent vector field to a family of geodesics, $D$ is the field with integral curves parametrising the family, and $T$ and $R$ are the torsion and curvature. It is obvious that torsion and curvature have equal status here, and there is no fundamental justification for neglecting either. Like spacetime curvature, spacetime torsion describes the geometry \emph{directly}, without needing the mediation of field equations and coupling constants, as matter fields do. There are of course systems such that torsion is small and can be neglected, just as there are systems in which Newtonian gravitation is an adequate approximation to General Relativity. It would be risky to assume that Inflationary cosmology belongs to either category.

We have seen however that, if we introduce torsion into cosmology, then there is a price to be paid$\,$: in all but a few special cases, Inflation no longer \emph{completely} isotropises the early Universe. But recent observations show that this may be precisely what we need: the evidence for anisotropy is hard to ignore \cite{kn:eoin}. If one or more of the various kinds of cosmic anisotropy should be confirmed observationally, then torsion may be a way ---$\,$ or \emph{the} way ---$\,$ to reconcile such observations with Inflation.

However, the operative word here is, as it should be, ``\emph{may}$\,$''. We have seen that torsion can, extremely naturally, account for \emph{some} forms of anisotropy, but by no means all. Anisotropy (at the end of Inflation) along more than one axis, or along one axis but not minimal there$\,$; anisotropy that involves no rotation, or is such that the amount of anisotropy is not correlated with the amount of rotation in the precise way described above$\,$; anisotropy on extremely large scales, corresponding to inflationary modes which left the horizon early in the inflationary era$\,$: any of these, if confirmed, might present a serious challenge to our proposed reconciliation of Inflation with anisotropy.

We feel that the last of these possibilities is the most serious objection to torsional Inflation. We will show elsewhere how it can be addressed.

\addtocounter{section}{1}
\section*{\large{\textsf{Acknowledgement}}}
The author is deeply grateful to Dr. Soon Wanmei for all of her many helpful comments.

\end{document}